\documentclass[twocolumn,showpacs,preprintnumbers,amsmath,amssymb]{revtex4}

\usepackage{graphicx}
\usepackage{dcolumn}

\begin{document}

\title{A probability-conserving dissipative Schr\"odinger equation}

\author{Michel van Veenendaal, Jun Chang, and A. J. Fedro}

\affiliation{%
Dept. of Physics, Northern Illinois University,
De Kalb, Illinois 60115\\ 
Advanced Photon Source,
  Argonne National Laboratory, 9700 South Cass Avenue, Argonne,
Illinois 60439}%

\date{\today}

\begin{abstract}
Dissipative effects on a microscopic level are included in the Schr\"odinger equation. When the decay between different local levels as a result of the coupling to a bath, the Schr\"odinger equation no longer conserves energy, but the probability of the states is conserved. The procedure is illustrated with several examples that include direct electronic decay and damping of local phonons (vibrational levels). This method significantly reduces the calculational effort compared to conventional density matrix techniques.
\end{abstract}

\pacs{ 03.65.Yz, 05.70.Jk} 

\maketitle
{\it Introduction}$.-$ 
A very diverse range of research fields such as solar energy conversion, quantum information storage, photosynthesis, and light controlled devices depend on a deeper understanding of the nonequilibrium dynamics of the system under consideration.  The increased use of pump-probe type experiments, in particular with the coming on line of the  next-generation X-ray sources, provides increasingly detailed knowledge of the transient electronic properties of a wide variety of materials \cite{Chen}.
The conventional approach to nonequilibrium problems is the density matrix method \cite{Lin}. The  diagonal  elements $\rho_{ii}(t)$ of the density matrix describe the population or probability of state $i$ as a function of time $t$, whereas the off-diagonal matrix elements $\rho_{ii'}(t)$ with $i'\ne i$ give the phase or coherence of the system.  This method is a powerful approach when one wants to obtain a detailed understanding of the nonequilibrium dynamics. However, the density matrix method is not entirely without problems. Within certain approximation schemes, the density matrix may become diverging \cite{Palmieri}. This problem can be avoided by the use of Lindblad operators \cite{Lindblad}, whose implementation is often far from straightforward, even for simple systems. Although this can in principle be solved, a more serious disadvantage of the density-matrix technique is the difficulty in handling large systems. In many problems, the size of the basis $N$ needed to describe the local system  can easily be of the order $N\sim 10^{2-6}$. Within a density matrix framework, this requires the evaluation of $N^2$ density matrix elements, which is intractable even for small $N$ and impossible for large $N$. In contrast, if one could directly study the nonequilibrium dynamics of the  wavefunction $|\psi(t)\rangle=\sum_{i=1}^N c_i(t) |\psi_i\rangle$, where the $|\psi_i\rangle $ form the basis of the system under consideration in bra-ket notation, only $N$ coefficients need to be calculated. The density matrices can then be directly obtained through $\rho_{ij}(t)=c_i^*(t)c_j(t)$. 

The goal of this Letter is to  describe dissipative processes  with an effective probability-conserving Schr\"odinger equation, 
\begin{eqnarray}
i\hbar \frac{d|\psi(t)\rangle}{dt} =\left (H_0+iD \right )| \psi(t)\rangle,
\label{SE}
\end{eqnarray}
where  $H_0$ is the Hamiltonian of the system and $D$ describes the effective dissipation. Several phenomenological expressions for $D$ have been suggested, such as $D\sim (1/\hbar) \ln (\psi/\psi^*)$ \cite{Kostin} and $D\sim (\partial/\partial t) \ln (\psi^*\psi)$ \cite{Davidson}. However, these damping terms primarily dealt with the quantum behavior of macroscopic variables, such as a Ginzburg-Landau type wavefunction or the phase difference across a Josephson tunnel junction. Our interest lies in incorporating microscopic dissipative mechanisms within systems consisting of, for example, an  ion (e.g. transition metal or rare earth) surrounded by ligands. This "local" system is considered part of a larger system such as a molecule in solution or a solid. The latter constitute the effective surroundings that can dissipate energy from the local system. First, we describe the nature of the dissipative Schr\"odinger equation in the presence of a microscopic dissipation mechanism. Our focus lies on ultrafast intersystem couplings and we only consider the zero-temperature formalism. We then demonstrate how to incorporate electronic and vibronic dissipation.

{\it Dissipative Schr\"odinger equation}$.-$
 The Schr\"odinger equation in the presence of a bath is given by
\begin{eqnarray}
i\hbar \frac{d|\psi(t)\rangle}{dt} =(H_0+H_B) |\psi(t)\rangle,
\end{eqnarray}
where $H_0$ is the Hamiltonian of the local system and $H_B$ the Hamiltonian of the bath plus the interaction of the local  system with the bath.  
When writing the coefficient $c_i$ for a particular vector in the basis in terms of an amplitude $a_i(t)=|c_i(t)|$ and a phase $\varphi_i$, or $c_i(t)=a_i(t) e^{i\varphi_i(t)}$, we can write for the change in coefficient due to the presence of the bath
\begin{eqnarray}
\left . \frac{dc_i(t)}{dt}\right |_B =
\left . \frac{da_i(t)}{dt}\right |_B  e^{i\varphi(t)}
+i \left . a_i(t) e^{i\varphi(t)} \frac{d\varphi_i(t)}{dt}\right |_B. 
\end{eqnarray}
The latter terms gives the change in phase, which causes an embedding of the local system in its surroundings. Since the bath has many degrees of freedom, we assume that the phase change of the local system due to the bath can be neglected based on law of large numbers. In addition, due to the complexity of the surroundings, the precise nature of this embedding is often very difficult and can usually only be taken into account in some effective way. We therefore  consider only the strength of the decay. Now let us assume for the moment that we are able to determine an expression for the change in amplitude $P_i=a_i^2$ in a particular basis (e.g in the absence of intersystem coupling constants)
\begin{eqnarray}
\left . \frac{dP_i}{dt}\right |_B
=2a_i \left . \frac{d a_i}{dt}\right |_B
=f(\{ P_i \} ),
\label{rate}
\end{eqnarray}
where $f$ is a function of the probabilities $P_i$. In the following, we demonstrate how to derive differential equations for $P_i$ and consider the effects of their incorporation.
The change in the coefficient due to the bath is then given by
\begin{eqnarray}
\left . \frac{dc_i}{dt}\right |_B 
=\frac{1}{2a_i}\left . \frac{dP_i}{dt}\right |_B  e^{i\varphi}  
=\left . \frac{1}{2}\frac{d\ln P_i}{dt}\right |_B  c_i.
\end{eqnarray}
This leads to a dissipative term in Eq. (\ref{SE}) given by
\begin{eqnarray}
D=\frac{\hbar}{2}\sum_i \frac{d\ln P_i(t)}{dt} |\psi_i \rangle \langle \psi_i | .
\label{D}
\end{eqnarray}
The dissipation does not necessarily have to be diagonal. After deriving the diagonal dissipation in a particular basis, a unitary transformation to a different (more suitable) basis can be made. Note that if Eq. (\ref{rate}) conserves the total probability, the probability is also conserved in the dissipative Schr\"odinger equation.

{\it Electronic transitions}$.-$
In the following, we give several examples that demonstrate the use of the dissipative Schr\"odinger equation. First, we consider electronic transitions, where a decay mechanism can be given that directly couples the states of the local system. This decay mechanism does not necessarily conserve the number of electrons of the local system, i.e. the problem can also be classified as an  "open quantum system". A typical example is the Fano problem \cite{Fano}, where a local state is coupled to a continuum and the particle number at the local state is not conserved. However, alternatively, one can also view open quantum problems as  closed systems when considering the probability of the local states, which is is conserved. In the Fano problem, there are two states: $|1\rangle$ and $|0\rangle$ with and without a particle on the local site. The total probability for the local states is conserved and governed by the equations
\begin{eqnarray}
\frac{dP_1(t)}{dt}&=&-2\Gamma P_1(t) ~~~ {\rm and}~~~\frac{dP_0(t)}{dt}=2\Gamma P_1(t).
\end{eqnarray}
The corresponding dissipative Schr\"odinger equation gives
\begin{eqnarray}
\frac{dc_1(t)}{dt}&=&-\Gamma c_1(t) ~~~ {\rm and}~~~\frac{dc_0(t)}{dt}=\Gamma \frac{c_1^2(t)}{c_0(t)}.
\end{eqnarray}
The solution is straightforward and gives, as expected, $c_1=e^{-\Gamma t}$ and $c_0=\sqrt{1-e^{-2\Gamma t}}$.
 
\begin{figure}[t]
\includegraphics[width=0.9\columnwidth]{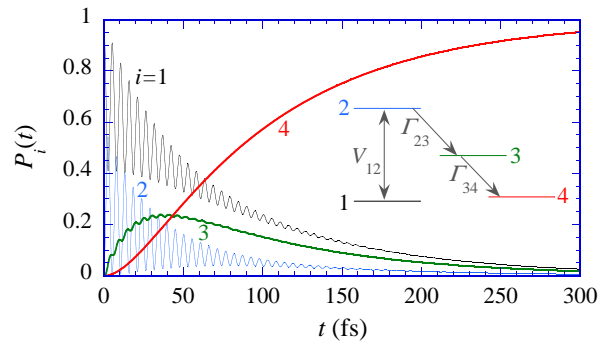}
\caption{(color online) The time dependence of the probabilities of a four-level system, with hybridization between states 1 and 2 and a cascade from $2\rightarrow 3 \rightarrow 4$, see inset. $\langle 1|H_0|2\rangle =0.3$ eV and $\langle 2|H_0|2\rangle -\langle 1|H_0|1\rangle =0.5$ eV. $dP_2/dt=-2\Gamma_{23} P_2$, $dP_3/dt=-2\Gamma_{34}P_3+2\Gamma_{23} P_2$, and  $dP_4/dt=2\Gamma_{34} P_3$  with $\hbar\Gamma_{23}=20$ meV (corresponding to a decay time of 66 fs), $\hbar\Gamma_{34}=10$ meV (33 fs).}
\label{level4} 
\end{figure}

Although the result for the Fano-type problem is hardly surprising, the real advantage lies in  situations where both electronic coupling and decay are present. Figure \ref{level4} shows an example with four levels, where the details are given in the caption. The initial state 1 is coupled via hybridization to level 2, which is higher in energy (the parameters are given in the caption). State 2 then shows a cascade decay via level 3 into level 4. State 1 and 2 show the typical quantum oscillations associated with hopping between two levels. Some oscillations are still observed in state 3 due to its coupling to state 2. The probability of finding state 4 steadily increases as that of initial state 1 decreases through intermediate states 2 and 3.

{\it Phonon decay}$.-$
In many dissipative processes, the number of particles in the local system is conserved and energy is lost through the damping of atomic vibrations. This situation occurs, for example, when two states with different interactions with the surroundings are coupled. Following an intersystem crossing, phonon modes (corresponding to, for example, local vibronic oscillations) are excited that decay on the femtosecond scale through, e.g. intramolecular energy redistributions. Let us first consider the decay of the excited phonon states and then incorporate this dissipative mechanism into a model for intersystem crossing.
\begin{figure}[t]
\includegraphics[width=0.9\columnwidth]{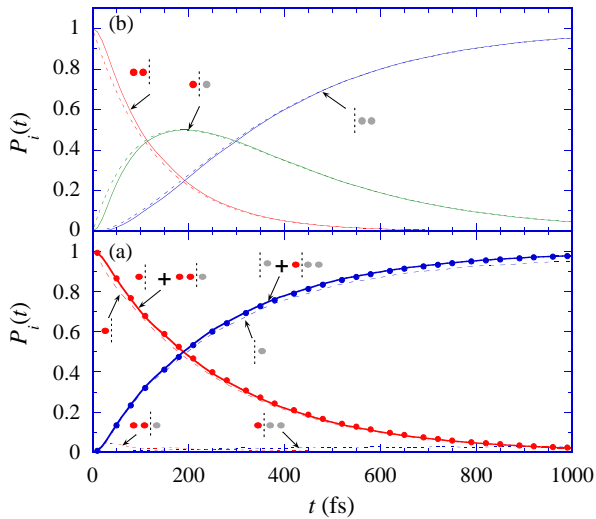}
\caption{(color online) Exact solutions of phonon damping for a small number of excited phonons. (a) Damping for an initial state with $n=1$ phonon. The red lines correspond to the probability of the effective $n_{\rm eff}=1$ consisting of $a^\dagger|0\rangle$ (dashed) which couples to the $a^\dagger a^\dagger b_k^\dagger|0\rangle$ states (dotted gives the total probability). The solid line gives the sum of both probabilities. The blue lines correspond to the effective $n_{\rm eff}=0$ states consisting of the $b_k^\dagger|0\rangle$ (dashed) and the $a^\dagger b_k^\dagger b_{k'}^\dagger|0\rangle$ (dotted) states. The solid blue line gives the sum of both probabilities. The thick dots correspond to the rotating-wave approximation (only the $a^\dagger|0\rangle$ and $b_k^\dagger|0\rangle$ are included). Note that in the limit $\Gamma\ll \hbar \omega$, the results are almost identical. (b) Phonon damping for $n=2$ in the rotating wave approximation. The probabilities for $n=2$ (red), 1 (green), and 0 (blue) are shown. The dashed lines give the analytical results.   }
\label{dampedHO} 
\end{figure}
\noindent

Let us take a Hamiltonian that couples a local vibrational mode to a bath of phonon modes
\begin{eqnarray}
H_B&=&\hbar \omega a^\dagger a + \sum_k [\hbar \omega_k b_k^\dagger b_k+ V_k (ab_k^\dagger +a^\dagger b_k^\dagger+ {\rm H.c.}) ],\nonumber
\end{eqnarray}
where $a^\dagger$ and $b_k^\dagger$ are the creation operators for an excitation in the local and continuum modes of energy $\hbar\omega$ and $\hbar\omega_k$, respectively. The last term with $V_k$ describes the quadratic coupling to the bath. We want to obtain an equation for the probabilities $P_n$ of having  $n$ excited phonons.  The fastest damping processes are primarily due to intramolecular energy redistributions, which are generally almost independent of temperature. We therefore take the limit of zero temperature. The damping due to $H_B$ can be solved numerically, but only for a very limited number of excited levels. Figure \ref{dampedHO} shows the case for $n=1$ for a local vibronic mode with $\hbar\omega=30$ meV and a continuum of bath levels from 0 to $3\hbar \omega$; the coupling $V_k=6$ meV. The contribution of the terms that do not conserve the number of phonons, $a^\dagger b_k^\dagger$ and $ab_k$, is relatively small since these excitations have an energy of at least $\hbar\omega$. Particle conserving terms can occur at the same energy. Furthermore, the non-phonon conserving terms act as an effective embedding. The effective boson $\alpha |1\rangle+\sum_k \beta_k |2,k\rangle$, where $|n,k\rangle$ is a state with $n$ excitations of the local vibronic mode and a boson with index $k$ in the continuum, decays to $\alpha' |0,k\rangle+\sum_{kk'} \beta'_{kk'} |1,k\rangle$ in an almost identical fashion as $|1\rangle$ decays to $|0\rangle$, see Fig. \ref{dampedHO}. This is only correct in the limit that the broadening of the local vibronic level due to the continuum is less than $\hbar\omega$. However, since an intersystem crossing on a timescale of 100 fs corresponds to an effective broadening of 3 meV, this condition is generally satisfied. In the remainder, we assume that the ultrafast decay can be described by (effective) bosons that couple to the continuum in a particle-conserving manner. This is also known as the rotating-wave approximation. 
We want to study the equation of motions for a system with $n$ excited local vibronic modes $A^\dagger_n|0\rangle$, where $A^\dagger=(a^\dagger)^n/\sqrt{n!}$ and $|0\rangle$ is the vacuum state. The time dependence of $A^\dagger_n(t)=e^{-iHt/\hbar}A^\dagger_n e^{iHt/\hbar}$ is given by 
\begin{eqnarray}
\frac{dA^\dagger_n(t)}{dt}=-i n\omega A^\dagger_n(t)-i\sqrt{n} \sum_k V_k b_k^\dagger (t) A_{n-1}^\dagger (t).
\label{An}
\end{eqnarray}
A similar equation can be obtained for $b_k^\dagger(t)$. Using  a Markov-type approximation, 
$\sum_k V_k^2 \int_0^t dt' e^{\pm i\omega_k (t-t')} =\Gamma \delta (t-t')$,
 the solution can be obtained
\begin{eqnarray}
A_n^\dagger(t)&=&e^{-inzt} A_n^\dagger \\ &&-i e^{-inzt}\sqrt{n} \int_0^t dt' \sum_k V_k e^{i(nz-\omega_k)t'} 
b_k^\dagger A_{n-1}^\dagger (t') , \nonumber
\end{eqnarray}
with the complex frequency $z=\omega-i\Gamma$.
For $N=1$, the problem is equivalent to a Fano-type decay and we can write $A_1^\dagger(t)=[c_1(t)a^\dagger+\sum_k c_k(t) b_k^\dagger]|0\rangle$. The probabilities are given by $p_1(t)=|c_1(t)|^2=e^{-2\Gamma t}$ and $p_0(t)=\sum_k |c_k(t)|^2=1-e^{-2\Gamma t}$.  After obtaining the result for $N=1$, we can derive general expression for arbitrary $N$ via iteration. This problem is separable and the solution, after a somewhat lengthy but straightforward derivation, is given by $A_N^\dagger(t)=[c_1(t)a^\dagger+\sum_k c_k(t) b_k^\dagger]^N|0\rangle$. The probability of finding $n$ local phonon modes at a particular time for an initial condition of $N$ phonon,
$P_n (t)=(N!/(N-n)! n!) p_1^n p_0^{N-n}$. In general, an equation can be derived for the probability of $n$ phonons
\begin{eqnarray}
\frac{dP_n(t)}{dt}=-2n\Gamma P_n(t)+2 (n+1)\Gamma P_{n+1} (t),
\label{DEPn}
\end{eqnarray}
which is the microscopic decay equation \cite{Bopp} that can be incorporated in the dissipative Schr\"odinger equation.
\begin{figure}[t]
\includegraphics[width=0.9\columnwidth]{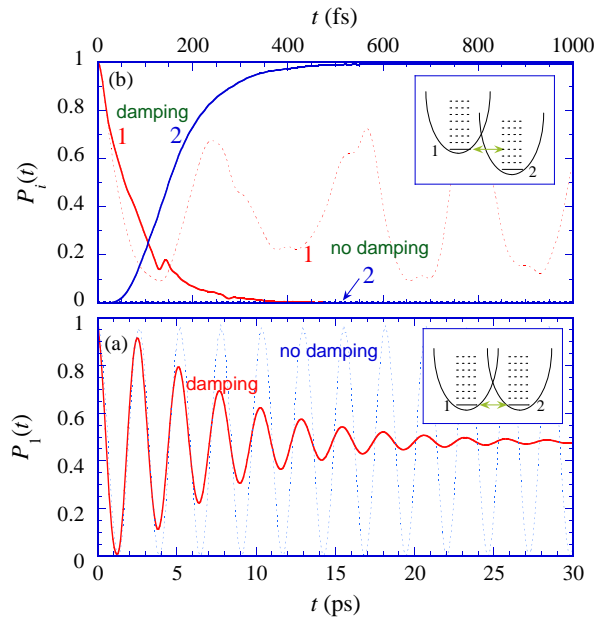}
\caption{(color online) (a) Occupation $P_1$ of the lowest vibrational level ($n=0$) of state 1 as a function of time. There is no bias between states 1 and 2 ($\Delta=0$), see inset. The initial state is state 1 with $n=0$. The dotted blue and solid red lines give the result in the absence and presence of damping, respectively. (b) The occupations $P_i$ of the lowest vibrational levels of states 1 (red) and 2 (blue) in for a bias equal to the phonon self-energy ($\Delta=\varepsilon$), see inset. The dotted and solid  lines are in the absence and presence of damping, respectively. Note the different timescales for (a) and (b). The parameters are $\varepsilon=0.2$ eV, $\hbar\omega=$30 meV, $V=$20 meV, $\hbar\Gamma=10$ meV.}
\label{Pph} 
\end{figure}
 
We now include the decay in, e.g. the  Hamiltonian \cite{Jaynes}
\begin{eqnarray}
H&=&\hbar \omega a^\dagger a +\sum_{i=1,2}[
(E_i +\sqrt{\varepsilon_i \hbar \omega}(a+a^\dagger)n_i ]  \nonumber \\
&& +V(c_1^\dagger c_2+c_2^\dagger c_1) ,
\end{eqnarray}
which describes two states at energy $E_i$ that couple differently to a phonon (e.g. a local vibronic mode) of energy $\hbar\omega$ via the coupling $\sqrt{\varepsilon_i \hbar \omega}$; the states couple with each other with a strength $V$. Since states 1 and 2 have different equilibrium positions, a transition between between induces an oscillation of the vibronic mode. Using the dissipative Schr\"odinger in Eqs. (\ref{SE}) and (\ref{D}) in combination with the change in phonon occupations as given by Eq. (\ref{DEPn}), we can effectively damp these oscillations. However, we need to make a choice in what basis to define the decay. In the limit $V=0$, the Hamiltonian separates into two independent displaced harmonic oscillators. Defining the occupations $P_{n}$ in the displaced bases leads to the physically intuitive situation that, in the absence of $V$, each state relaxes to its equilibrium position. Since, only the relative equilibrium positions are of importance, we can take $\varepsilon_1=0$ and $\varepsilon_2=\varepsilon$. Accounting for the phonon self-energy $\varepsilon_i$, the bias is then given by $\Delta=E_1-E_2+\varepsilon$. Figure \ref{Pph} demonstrates the effects of including the damping of the phonon oscillations on the occupations. The parameters are given in the caption.

In Fig. \ref{Pph}(a), the time-dependence occupation of the lowest vibrational level ($n=0$) of state 1 in the absence of a bias ($\Delta=0$, see inset in Fig. \ref{Pph}(a)) is given. This state is also the initial wavefunction. In the absence of damping of the local mode, there is an oscillation between the $n=0$ levels of state 1 and 2. Since the direct coupling between $n=0$ levels of state 1 and 2 is $Ve^{-\varepsilon/\hbar\omega}$ is very small, the effective coupling between the two levels occurs through the states where the Franck-Condon factors are large. This effectively creates an energy barrier with a height of the order of $\varepsilon$ between the two states.  The inclusion of damping causes a decoherence of the oscillations that damp out \cite{Budini}. Figure \ref{Pph}(b) shows the results in the presence of a bias ($\Delta=\varepsilon$), see inset. For this value of the bias, the coupling is relatively strong since the lowest vibrational level of state 1 is at the maximum of the Franck-Condon continuum \cite{MvV}. In the absence of dissipation, there is a periodic recurrence of occupation in state 1, related to the oscillation period of the local phonon. The occupation of the lowest vibrational level of state 2 is negligibly small in this case. When damping of the local phonon mode is introduced, this recurrence is strongly reduced \cite{Kuhn}. In addition, there is a steady increase in occupation of $n=0$ of state 2, which becomes close to 100\% occupied on the timescale of a few 100 fs.

{\it Conclusions}.$-$ We have demonstrated how dissipative effects can be included on a microscopic level in a Schr\"odinger equation by incorporating differential equations that describe the change in microscopic occupations. The results are intuitive and relatively straightforward to incorporate. If the decay equations are properly defined, the Schr\"odinger equation does not conserve energy, although probability is still conserved. The procedure was illustrated with examples that included a direct electronic decay and a damping of local phonon levels. The described method makes the size of the calculational efforts proportional to the size of the basis set $N$, which is a strong reduction to the commonly-used density matrix method (proportional to $N^2$). This makes this approach extremely useful in describing relatively large systems that undergo ultrafast decay of an excited state. Future research should involve the inclusion of the effects of temperature.

{\it Acknowledgments}.$-$ 
This work was  supported by  the U.S. Department of Energy (DOE), DE-FG02-03ER46097, and NIU's Institute
for Nanoscience, Engineering, and Technology under a grant from the U.S. Department of Education. Work at
Argonne National Laboratory was supported by the U.S. DOE, Office of Science, Office of Basic Energy Sciences, under contract DE-AC02-06CH11357.

\end{document}